\pdfoutput=1
\documentclass{article}
\usepackage{arxiv}
\usepackage[utf8]{inputenc}
\usepackage{amsmath,amssymb}
\usepackage{mathtools}
\usepackage{physics}
\usepackage{listings}
\usepackage{xspace}
\usepackage{xcolor}
\usepackage{ulem}
\usepackage[square,sort,comma,numbers]{natbib}

\DeclareUnicodeCharacter{2207}{$\nabla$}
\DeclareUnicodeCharacter{22C5}{$\cdot$}
\DeclareUnicodeCharacter{03B3}{$\gamma$}
\DeclareUnicodeCharacter{03BA}{$\kappa$}
\DeclareUnicodeCharacter{03BC}{$\mu$}
\DeclareUnicodeCharacter{03C6}{$\phi$}

\newcommand{\python}{\textbf{\textsf{Python}}\xspace}
\newcommand{\sympy}{\textbf{\textsf{SymPy}}\xspace}
\newcommand{\openfoam}{\textbf{\textsf{OpenFOAM}}\xspace}

\title{A multi-physics compiler for generating numerical solvers from differential equations}

\author{ \hspace{1mm}John T. Maxwell III, Morad Behandish\thanks{Corresponding author \\\textit{Email address:} \texttt{morad.behandish@sri.com} (Morad Behandish)}\ , and S{\o}ren Taverniers\\
\\
	Palo Alto Research Center (PARC), a part of SRI International, 3333 Coyote Hill Road, Palo Alto, CA 94304, USA\\
}

\renewcommand{\shorttitle}{\textit{arXiv} Template}

\begin{document}
\maketitle

\section{Introduction}

\subsection{Motivation}

We develop a tool that enables domain experts to quickly generate numerical solvers for emerging multi-physics phenomena starting from a high-level description based on ordinary/partial differential equations and their initial and boundary conditions over a symbolic spacetime domain. This ``multi-physics'' compiler aims to bridge the gap between problem formulation and computation, which historically has spanned years or even decades. Specialized numerical solvers in areas such as computational fluid dynamics (CFD) often present a barrier to novice end users not well-versed in the intricacies of their underlying schemes, and requiring surgical modifications when coupling with additional physical components initially not accounted for by the solver. Through the use of an intermediate language that is neutral between classical and exterior calculus, the compiler generates correct-by-construction numerical source code that offers guarantees of immutable physical principles like conservation laws at the discrete level. We present a proof of concept for the multi-physics compiler through some examples involving compilation to \openfoam \cite{Jasak2007}. A specific use case that the compiler is well-suited for involves equation modification approaches, where the aim is to use simple numerical schemes such as central differencing through the additional of artificial terms to the original governing equations of the multi-physics problem \cite{Mirjalili2021,Mirjalili2022,Mirjalili2023}.

\subsection{Paper Outline}

The paper proceeds as follows. Section \ref{section_templates} introduces an intermediate representation called ``grid templates'' and explains how differential equations get expanded into these constructs.  Section \ref{section_stencils} introduces grid stencils as grid templates with added information on the geometry of the computational grid. Section \ref{section_OpenFOAM} shows an example of grid templates in action when compiling to numerical source code readable by \openfoam. Finally, the appendices shows how a simple system of differential equations being specified in \python, and the C files that the compiler produces for those equations.

\section{Grid Templates}
\label{section_templates}

Converting differential equations (DEs) to a numerical solver is not straightforward since there is some ambiguity about how these equations should be discretized. We want the user to be able to control this ambiguity in a platform-independent way, i.e., without committing to a specific computational framework.  Moreover, in addition to classical DEs, we also like to be able to compile exterior DEs. We therefore introduce an intermediate representation that we call {\it grid templates}. Grid templates provide a discrete representation of DEs that is neutral between classical and exterior calculus, and facilitate disambiguation between various discretization options available to the user. The compiler provides a set of default templates for classical and exterior calculus operators, but the user can override these default heuristics on a variable-by-variable basis if necessary.

An example of a grid template expanding a temperature gradient, $\gradient T$, is given in (\ref{eq_sample_template}),
\begin{multline}
\frac{\operatorname{sum}(\operatorname{in}(f, \partial(\operatorname{dual}(c))), T(\operatorname{centroid}(f)) \operatorname{normal}(f) \operatorname{orientation}(f, \operatorname{dual}(c)))}{\operatorname{size}(\operatorname{dual}(c))}.
\label{eq_sample_template}
\end{multline}

We will provide a detailed description of this template in Section \ref{section_gradient}.

\subsection{Grid Geometry Elements}

The grid template notation represents geometry elements using typed variables. There are currently four types of element variables: volumes (v), volume centers (c), volume faces (f), and the lines between two volume centers (l). This set can be extended to include volume edges, volume vertices, volumes centered on vertices, and the faces of volumes centered on vertices.

\subsection{Grid Time Elements}

Time is represented in grid notation as a sequence of discrete instants.  We use $t$ as the time variable, such that $t+1$ represents the next time instant, and $t-1$ represents the previous time instant.

\subsection{Grid Geometry Operators}

The grid template notation supports the following geometric functions, inspired by the concept of dual cell complexes from algebraic topology \cite{Hirani2003}.

\begin{itemize}

\item $\operatorname{boundaries}(l)$ - the end points of a line (printed as $\partial(l)$)
\item $\operatorname{boundaries}(v)$ - the faces of a volume (printed as $\partial(v)$)
\item $\operatorname{centroid}(f)$ - the centroid of a face
\item $\operatorname{dual}(c)$ - the volume containing a center
\item $\operatorname{dual}(f)$ - the line between two centers that passes through a face
\item $\operatorname{dual}(l)$ - the face that a line passes through
\item $\operatorname{dual}(v)$ - the center of a volume
\item $\operatorname{normal}(f)$ - the normal of a face
\item $\operatorname{normal}(l)$ - the line as a vector
\item $\operatorname{orientation}(c, l)$ - the orientation of an end point relative to its line ($\pm$ 1)
\item $\operatorname{orientation}(f, v)$ - the orientation of a face relative to its volume ($\pm$ 1)
\item $\operatorname{size}(f)$ - the area of a face
\item $\operatorname{size}(l)$ - the length of a line
\item $\operatorname{size}(v)$ - the volume of a volume

\end{itemize}

In our experience so far, boundaries(), centroid(), dual(), normal(), orientation(), and size() are sufficient to give us enough flexibility for defining grid templates.

\subsection{Physical Variables and Constants}

Physical variables take two arguments: the geometric element that they apply to and the time instant that they apply to.  For instance, $T(c,t)$ represents temperature at the given center variable and the given time variable. Likewise, $T(\operatorname{centroid}(f), t+1)$ represents temperature at the centroid of the given face variable and the next time instant after the given time variable. By convention, if the time argument is not given (e.g. $T(c)$), then it is implicitly assumed to be $t$.

Physical variables can also have properties, such as their value type (i.e., scalar, vector, or tensor), their meaning (e.g., ``temperature''), their range (e.g., ``positive''), and how they are initialized. Physical constants are like physical variables except for the fact that they do not take a geometric argument or a time argument.

\subsection{Physical Functions}

The grid template notation is based on \sympy \cite{Meurer2017}, and uses \sympy functions where possible. For instance, +, *, -, $\curl$, $\gradient$, $\divergence$, and $\cdot$ (dot product) are implemented using \sympy functions.  Other functions are implemented as \sympy unknown functions.

\subsection{Expansion}

To compile a differential equation to a computational platform, we first expand the terms of the DEs using grid templates. In order to do the expansion, we need to know what geometric element each variable applies to. Our default assumption is that variables are defined at centers, because most CFD sofware packages (including \openfoam) use collocated grids where variables are stored at cell centers, and our initial testbed for the compiler uses \openfoam as the target language. For instance, the heat equation $\divergence \gradient T = 0$ is implemented as $\divergence \gradient T(c) = 0$.

\subsection{Gradient Templates}
\label{section_gradient}

If the user has not provided an explicit template for a gradient like $\gradient T(c)$, then the compiler will create the default template
\begin{multline}
\frac{\operatorname{sum}(\operatorname{in}(f, \partial(\operatorname{dual}(c))), T(\operatorname{centroid}(f)) \operatorname{normal}(f) \operatorname{orientation}(f, \operatorname{dual}(c)))}{\operatorname{size}(\operatorname{dual}(c))}
\label{eq_default_gradient}
\end{multline}
The first part of \eqref{eq_default_gradient}, $\operatorname{sum}(\operatorname{in}(f, \partial(\operatorname{dual}(c))), x(f,c))$, where

\begin{align}
x(f,c)=T(\operatorname{centroid}(f)) \operatorname{normal}(f) \operatorname{orientation}(f, \operatorname{dual}(c)),
\label{expression}
\end{align}
takes a sum over the faces of the volume that is the dual of $c$.  Mathematically it is equivalent to $\sum_{f \in \partial(\operatorname{dual}(c))}{x(f,c)}$.  Note that $f$ is scoped by $\operatorname{sum}$, so that if another variable named $f$ is introduced by another $\operatorname{sum}$ inside of $x(f,c)$, the embedded variable takes precedence within its scope.

The second part of \eqref{eq_default_gradient} is a product of terms that indicates what should be summed: $T(\operatorname{centroid}(f))$ is the value of $T$ at the centroid of the face $f$ introduced by the sum,
$\operatorname{orientation}(f, \operatorname{dual}(c))$ is the orientation of the face $f$ relative to its volume ($\operatorname{dual}(c)$), and $\operatorname{normal}(f))$ is the normal vector of the face $f$ scaled by the area of $f$. Each face has a globally fixed orientation. Its orientation relative to one of its volumes will be positive and its orientation relative to the adjacent volume will be negative. This means that the flux of a face will be positive for one volume and negative for the adjacent volume, as desired.

The final part of \eqref{eq_default_gradient} is the denominator, $\operatorname{size}(\operatorname{dual}(c))$.  This normalizes the sum given by the first and second part by dividing it by the volume of the cell containing $c$.  The normalization is to get $\gradient T$ at $c$, the center of the volume.

To summarize, this gradient template says that $\gradient T(c)$ is implemented on a grid by taking the value of $T$ at the centroid of each face, multiplying it by the normal of the face to get a vector, summing all of the vectors to get the overall gradient, and then dividing by the volume to get the gradient at the cell center.  We will discuss how to compute $T(\operatorname{centroid}(f))$ in the section on interpolation below (Section \ref{section_interpolation}).

\subsection{Divergence Templates}

If the user has not provided an explicit template for a divergence such as $\divergence \mathbf{U}(c)$,  then the compiler will create the default template
\begin{multline}
\frac{\operatorname{sum}(\operatorname{in}(f, \partial(\operatorname{dual}(c))), (\mathbf{U}(\operatorname{centroid}(f)) \cdot \operatorname{normal}(f) \operatorname{orientation}(f, \operatorname{dual}(c))))}{\operatorname{size}(\operatorname{dual}(c))}.
\label{eq_default_divergence}
\end{multline}
This template sums over face terms like the gradient template, but the face terms are the dot product of the oriented normal of the face and the divergence variable at the face instead of their multiplication.  Finally, the sum is divided by the size of the volume that contains $c$ in order the get the divergence at the center of the volume.

\subsection{Combining Templates}

When an operator is applied to an operator (e.g. $\divergence \gradient T(c)$), then the outer operator is expanded first.  This means that the Laplacian $\divergence \gradient T(c)$ is first expanded as
\begin{multline}
\frac{\operatorname{sum}(\operatorname{in}(f, \partial(\operatorname{dual}(c))), (\operatorname{normal}(f) \operatorname{orientation}(f, \operatorname{dual}(c)))\cdot \gradient T(\operatorname{centroid}(f)))}{\operatorname{size}(\operatorname{dual}(c))}.
\label{eq_default_laplacian}
\end{multline}
Next, $\gradient T(\operatorname{centroid}(f))$ is expanded using interpolation (see Section \ref{section_interpolation}) and substituted in.  This produces
\begin{multline}
\operatorname{sum}(\operatorname{in}(f, \partial(\operatorname{dual}(c))), \\ (\operatorname{normal}(f) \operatorname{orientation}(f, \operatorname{dual}(c)))\cdot (\operatorname{normal}(\operatorname{dual}(f)) \\ \operatorname{sum}(\operatorname{in}(c, \partial(\operatorname{dual}(f))), T(c) \operatorname{orientation}(c, \operatorname{dual}(f))) \\ /\operatorname{size}(\operatorname{dual}(f))^2))/\operatorname{size}(\operatorname{dual}(c)).
\label{eq_default_laplacian2}
\end{multline}

\subsection{Interpolation}
\label{section_interpolation}

In order the determine the value of a variable at the centroid of a face (e.g. $T(\operatorname{centroid(f)})$), we interpolate the values of the variable at the centers of the neighboring volumes. On a uniform grid, this becomes
\begin{multline}
\operatorname{sum}(\operatorname{in}(c, \partial(\operatorname{dual}(f))), T(c)/2).
\label{eq_default_interpolation}
\end{multline}
The dual of $f$ is the line between the centers of the neighboring volumes of $f$, denoted by $\partial(\operatorname{dual}(f))$). In other words, this template interpolates the variable, say $T$, at the face center by taking the average of its values at the centers of the neighboring volumes.

If we take the divergence of a product (e.g., $\div(\mathbf{U}(c) p(c))$), then we end up with a product of two variables at the face center (e.g., $\mathbf{U}(\operatorname{centroid}(f)) p(\operatorname{centroid}(f)$).  We could expand these variables separately and multiply the result, but we could also expand them together as
\begin{multline}
\operatorname{sum}(\operatorname{in}(c, \partial(\operatorname{dual}(f))), \mathbf{U}(c)p(c) /2).
\label{eq_default_interpolation2}
\end{multline}
This is one of the ambiguities that arise when trying to discretize continuous DEs. It is a nontrivial choice, since in some cases one of the expansions leads to a conservative discretization and the other one does not.

If we take the divergence of a gradient, then we end up with a gradient at the face center (e.g., $\gradient T(\operatorname{centroid}(f))$).  We could interpolate the gradient term like we did in \eqref{eq_default_interpolation}.  But we could also calculate the gradient directly by looking at the difference between $T(c)$ in the neighboring volumes,
\begin{multline}
\frac{\operatorname{sum}(\operatorname{in}(c, \partial(\operatorname{dual}(f))), T(c) \operatorname{orientation}(c, \operatorname{dual}(f))) \operatorname{normal}(\operatorname{dual}(f))}{\operatorname{size}(\operatorname{dual}(f))^2}.
\label{eq_default_interpolation3}
\end{multline}
In this template, the summation calculates the difference between $T$ at one end point of the line and $T$ at the other end point, where the line is between the centers of the neighboring volumes.  It is a difference because the orientations of the two end points have opposite signs.  The orientations of the end points are determined by the orientation of the line, and the orientation of the line is the same as the orientation of the face that is its dual.  So the difference is consistent with the orientation of the face.  The difference is divided by $\operatorname{size}(\operatorname{dual}(f))$ in order to determine the rate of change of $T$.  Finally, the rate of change is multiplied by the unit vector of the line ($\operatorname{normal}(\operatorname{dual}(f))/\operatorname{size}(\operatorname{dual}(f))$) to turn it into a gradient vector (the normal of a line is just the line itself).

Taking the gradient directly on the face is more efficient than calculating the gradient at the cell centers and interpolating the results to the faces, and often is also the better choice for generating conservative discretizations. Hence, by default the system takes the gradient directly on the face, but the user can override this default heuristic by defining a template according to \eqref{eq_default_interpolation},
\begin{multline}
\operatorname{sum}(\operatorname{in}(c, \partial(\operatorname{dual}(f))), \gradient T(c) /2).
\label{eq_default_interpolation4}
\end{multline}

\subsection{Hodge Star}

The Hodge star operator is a construct from exterior calculus. It moves a variable from one geometric element to its dual by multiplying by the size of the original geometric element and dividing by the size of its dual (note that size(c) is equal to 1). For instance, $\operatorname{Hodge}(T(v))$ approximates $T(v)$ in terms of $T(c)$ using
\begin{equation}
T(\operatorname{dual}(v)) \operatorname{size}(v).
\label{eq_Hodge1}
\end{equation}
Similarly, $\operatorname{Hodge}(T(l))$ approximates $T(l)$ in terms of $T(f)$ using
\begin{equation}
\frac{T(\operatorname{dual}(l)) \operatorname{size}(l)}{\operatorname{size}(\operatorname{dual}(l))}.
\label{eq_Hodge2}
\end{equation}

\subsection{Exterior Derivative}

The default expansion of the exterior derivative of a variable applied to a line ($d(T(l))$) is given by the template
\begin{equation}
\operatorname{sum}(\operatorname{in}(c, \partial(l)), T(c) \operatorname{orientation}(c, l)).
\label{eq_ext_deriv}
\end{equation}
The exterior derivative of a variable on a line is the difference between the variable's values at the end points.  The template calculates this by summing the end point values oriented relative to the line. Hence, one of orientations will be positive and the other one will be negative.  The orientations of the end points correspond to the orientation of the line, which is defined to be consistent with the orientation of the face that is the dual of the line.
The exterior derivative of a variable at a cell center ($d(T(c))$) is given by the template
\begin{equation}
\frac{\operatorname{sum}(\operatorname{in}(f, \partial(\operatorname{dual}(c))), T(f) \operatorname{orientation}(f, \operatorname{dual}(c)))}{\operatorname{size}(\operatorname{dual}(c))}.
\label{eq_ext_deriv2}
\end{equation}

\section{Grid Stencils}
\label{section_stencils}

Grid stencils are like grid templates except that they take some of the geometry of the grid into account.  Grid templates are agnostic to the grid geometry; they know that geometric elements have boundaries, but they do not know how many boundaries a geometric element has or how the boundaries are related to each other.  Grid stencils are grid templates that have been specialized to the geometry.

Consider the fully expanded grid template for $\gradient T(c)$,
\begin{multline}
\operatorname{sum}(\operatorname{in}(f, \partial(\operatorname{dual}(c))), \operatorname{normal}(f) \operatorname{orientation}(f, \operatorname{dual}(c)) \\ \operatorname{sum}(\operatorname{in}(c, \partial(\operatorname{dual}(f))), T(c)/2))/\operatorname{size}(\operatorname{dual}(c)).
\label{eq_gradient_template}
\end{multline}
If we specialize this to a uniform two-dimensional geometry, then we know that $\partial(\operatorname{dual}(c))$ has four faces.  For each face, we get two terms involving $T(c)$, one at the center of the volume and one at the center of the neighboring volume.  In each dimension, the opposing faces have opposite orientations.  This means that the terms at the center of the volume cancel each other and the remaining terms have opposite signs.  We group the terms for each dimension to get the vector
\begin{multline}
\left(\frac{T(i + 1,j) - T(i - 1,j)}{2 \Delta x}, \frac{T(i,j + 1) - T(i,j - 1)}{2 \Delta y}\right),
\label{eq_stencil}
\end{multline}
where the grid cell sizes $\Delta x$ and $\Delta y$ arise because $\operatorname{size}(\operatorname{dual}(c))$ is $\Delta x \Delta y$ in a uniform grid.

Grid stencils are useful when a computation needs to know a little more about the geometry than grid templates provide. For instance, while our current target language (\openfoam) has built-in discretization tools and only requires the use of grid templates to perform the compilation, other platforms may require source code that also incorporates the grid geometry.

\section{Compiling to \openfoam}
\label{section_OpenFOAM}

To compile a system of DEs to \openfoam, the compiler first expands the DEs using grid templates, as described above.  Then it translates the templates into an \openfoam-readable source code written in C. Next, the user compiles this source code and uses the resulting executable to simulate an \openfoam case. The problem geometry is completely stored in \openfoam case files, i.e., the compiler and the C code it generates are agnostic to the geometry. This means that the \openfoam compiler doesn't need grid stencils, it just needs grid templates.

Appendix A shows a sample user input in \python for compiling a system of DEs describing a simple two-phase flow. Appendices B, C and D show the output of the compiler comprised of the C source code for the solver, a C header file for initializing relevant fields, and a case file specifying numerical schemes for time stepping (when using built-in time-integration schemes) and discretization of spatial operators, respectively. The rest of this section describes how the compiler translates the grid templates derived from the user's input into the compiler's output.

\subsection{Patterns}

The compiler translates grid templates using specialized code, which involve default patterns that can be overridden by the user. For instance, \eqref{eq_pattern} shows a pattern for translating a simple interpolation,
\begin{multline}
\operatorname{sum}(\operatorname{in}(c, \partial(\operatorname{dual}(f))), \operatorname{?arg1}/2) \rightarrow \operatorname{interpolation}(\operatorname{?arg1}, \operatorname{linear}).
\label{eq_pattern}
\end{multline}
Patterns are applied to the grid templates bottom up. If the left part of a pattern matches, then it is replaced by the right part. Any symbol that begins with a question mark is a variable that can match any expression.

\subsection{Variables}

Template variables (e.g., $T(c)$) get translated into C variables.  Where possible, the C variables are given the same name as the template variable that they come from.  The compiler also introduces variables to represent internal computations when that makes things more clear. It tries to name these variables such that they can be easily understood by the user.

\subsection{Generation}

After the grid templates have been translated to \openfoam expressions using patterns or specialized code, the compiler generates C code from the \openfoam expressions.  For instance, ``interpolation(T, linear)'' is generated as ``fvc::interpolate(T)'' with an entry in the fvSchemes case file saying that the interpolation should be linear,

\begin{verbatim}
interpolationSchemes
{
    interpolate(T)    linear;
}
\end{verbatim}

\subsection{Initialization}

\openfoam variables are initialized in a createFields.H file that is used by the solver.  The user specifies that a variable should be initialized by reading in values from the \openfoam case by setting the variable's ``init'' property to T.  The compiler then generates

\begin{verbatim}
Info<< "Initializing temperature T\n" << endl;

volScalarField T
        (
            IOobject
            (
                "T",
                runTime.timeName(),
                mesh,
                IOobject::NO_READ,
                IOobject::AUTO_WRITE
            ),
            mesh
        );
\end{verbatim}
The user can also set the ``init'' property to an expression, such as $p / (R \rho)$.  The compiler will then generate code to initialize the variable with this value,

\begin{verbatim}
Info<< "Initializing temperature T\n" << endl;

volScalarField T
        (
            IOobject
            (
                "T",
                runTime.timeName(),
                mesh,
                IOobject::NO_READ,
                IOobject::AUTO_WRITE
            ),
            mesh,
            dimensionedScalar("T", p.dimensions() /
              (R.dimensions() * rho.dimensions()), scalar(0))
        );

T = p / (R * rho + dimensionedScalar("eps",
                   R.dimensions() * rho.dimensions(), 1e-10));
T.correctBoundaryConditions();
\end{verbatim}
If an expression includes a divisor that is known to be positive because of user declarations (such as R * rho above), the compiler will automatically add a small value ``epsilon'' to the divisor in order to avoid division-by-zero errors.

\subsection{Boundary Conditions}

The compiler automatically generates code to correct boundary conditions.  If a variable like $T$ is set to an expression, then the compiler will generate ``T.correctBoundaryConditions()'' immediately after it.  If a variable is solved for, then the generator will generate boundary correction code for the ``boundary\_correction'' property if it exists, and otherwise for the ``init'' property if it is an expression.  For instance, if  the ``init'' property for rhoU is $U * rho$, then the boundary correction code looks like

\begin{verbatim}
rhoU.boundaryFieldRef() == U.boundaryField() * rho.boundaryField();
\end{verbatim}

\subsection{Time Schemes}

The compiler currently supports all temporal integration schemes built into \openfoam (including forward/backward Euler and Crank-Nicolson) as well as the standard fourth-order Runge-Kutta method (RK4).  Rather than requiring that the user change the grid templates to reflect the desired time scheme, the compiler lets the user specify the desired time scheme directly and then the compiler generates the appropriate code for that time scheme.

For instance, if the user specifies that they want to use a built-in explicit time scheme, then the compiler will always use ``fvc::'' calls (explicit) rather than ``fvm::'' calls (implicit) for spatial operators.
If the user wants to use a built-in implicit time scheme, then the compiler will generate ``fvm::'' calls for spatial operators on variables that are being solved for and ``fvc::'' otherwise.
If the user specifies that they want to use the non-built-in scheme RK4, then the compiler will explicitly generate code to iterate over the various stages of the Runge-Kutta algorithm.

\section{Conclusion}

We have shown how differential equations (DEs) can be automatically compiled into numerical solvers that can be executed on computational frameworks like \openfoam by first expanding them into grid templates and then translating the grid templates into C code. The compiler encodes best practices for compiling DEs, but also allows the user to override its default heuristics when appropriate.  While so far we have demonstrated a proof of concept using compilation to C code readable by \openfoam, we plan to consider other target languages and computational frameworks in future work.

\bibliographystyle{unsrtnat}

\section{Appendix A: Sample User Input}
\label{section_input}
\begin{verbatim}
    # pressure
    gt = GridTemplates()
    center = gt.center_var()
    p_c = gt.variable("p", center, init = True, meaning = "pressure")
    gt.constraint(gt.divergence(gt.gradient(p_c)))

    # Darcy's law
    kappa = gt.constant("kappa", meaning = "permeability")
    mu = gt.constant("mu", meaning = "viscosity")
    u_c = gt.variable("u", center, value_type = gt.VECTOR, init = True,
                      meaning = "velocity")
    gt.definition(u_c, - kappa / mu * gt.gradient(p_c))

    # phase equation
    phi_c = gt.variable("phi", center, init = True, meaning = "phase",
                        value_type = gt.PERCENTAGE)
    n_c = gt.variable("n", center, value_type = gt.VECTOR)
    gt.definition(n_c, gt.normalized(gt.gradient(phi_c)))
    D = gt.constant("D", meaning = "diffusivity")
    gamma = gt.constant("gamma", meaning = "gamma")
    gt.constraint(gt.deriv(phi_c) + gt.divergence(u_c * phi_c), "=",
                  gt.divergence(D * gt.gradient(phi_c)) -
                  gt.divergence(gamma * phi_c * (1 - phi_c) * n_c))

    generator = OpenFoamGenerator(gt, "speciesFoam2", "temp")
    generator.generate_files()
\end{verbatim}

\section{Appendix B: Sample Solver}
\label{section_solver}
\begin{verbatim}
/*---------------------------------------------------------------------------*\
  =========                 |
  \\      /  F ield         | OpenFOAM: The Open Source CFD Toolbox
   \\    /   O peration     | Website:  https://openfoam.org
    \\  /    A nd           | Copyright (C) 2011-2018 OpenFOAM Foundation
     \\/     M anipulation  |
-------------------------------------------------------------------------------
License
    This file is part of OpenFOAM.

    OpenFOAM is free software: you can redistribute it and/or modify it
    under the terms of the GNU General Public License as published by
    the Free Software Foundation, either version 3 of the License, or
    (at your option) any later version.

    OpenFOAM is distributed in the hope that it will be useful, but WITHOUT
    ANY WARRANTY; without even the implied warranty of MERCHANTABILITY or
    FITNESS FOR A PARTICULAR PURPOSE.  See the GNU General Public License
    for more details.

    You should have received a copy of the GNU General Public License
    along with OpenFOAM.  If not, see <http://www.gnu.org/licenses/>.

Application
    speciesFoam2

\*---------------------------------------------------------------------------*/

#include "fvCFD.H"
#include "fvOptions.H"

int main(int argc, char *argv[])
{
    #define NO_CONTROL
    #include "setRootCaseLists.H"
    #include "createTime.H"
    #include "createMesh.H"
    #include "createFields.H"

    // ORIGINAL EQUATIONS
    // u(c) = -κ*∇(p(c))/μ
    // n(c) = normalized(∇(φ(c)))
    // ∇⋅(∇(p(c))) = 0
    // (d/dt φ(c)) + ∇⋅(φ(c)*u(c)) = ∇⋅(D*∇(φ(c))) - ∇⋅(γ*n(c)*φ(c)*(1 - φ(c)))

    // COMPILER PARAMETERS
    // time_scheme = implicit

    Info<< "\nStarting time loop for speciesFoam2\n" << endl;

    while (runTime.loop())
    {
        Info<< "Time = " << runTime.timeName() << nl << endl;

        surfaceScalarField flux1 = -kappa * mesh.magSf() * fvc::snGrad(p) / mu;
        volScalarField mag_grad_phi = mag(fvc::grad(phi));
        volVectorField n = fvc::grad(phi) / (mag_grad_phi +
             dimensionedScalar("eps", mag_grad_phi.dimensions(), 1e-10));
        volVectorField var2 = gamma * n * phi * (1 - phi);

        // Info<< "Processing ∇⋅(∇(p(c))) = 0" << endl;
        solve(
            fvm::laplacian(p)
        );

        // Info<< "Processing (d/dt φ(c)) + ∇⋅(φ(c)*u(c)) = ∇⋅(D*∇(φ(c))) - ∇⋅(γ*n(c)*φ(c)*(1 - φ(c)))" << endl;
        solve(
            fvm::ddt(phi) +
            fvm::div(flux1, phi)
            ==
            fvm::laplacian(D, phi) +
            -fvc::div(var2)
        );

        u = -kappa * fvc::grad(p) / mu;
        u.correctBoundaryConditions();

        p.correctBoundaryConditions();

        phi.correctBoundaryConditions();

        // writing time step on the disk
        runTime.write();

        Info<< "ExecutionTime = " << runTime.elapsedCpuTime() << " s"
            << "  ClockTime = " << runTime.elapsedClockTime() << " s"
            << nl << endl;
    }

    Info<< "End of speciesFoam2\n" << endl;

    return 0;
}
\end{verbatim}

\section{Appendix C: Sample createFields.H}
\label{section_createFields}
\begin{verbatim}
Info<< "Reading pressure p\n" << endl;

volScalarField p
        (
            IOobject
            (
                "p",
                runTime.timeName(),
                mesh,
                IOobject::MUST_READ,
                IOobject::AUTO_WRITE
            ),
            mesh
        );

Info<< "Reading phase phi\n" << endl;

volScalarField phi
        (
            IOobject
            (
                "phi",
                runTime.timeName(),
                mesh,
                IOobject::MUST_READ,
                IOobject::AUTO_WRITE
            ),
            mesh
        );

Info<< "Reading transportProperties\n" << endl;

IOdictionary transportProperties
        (
            IOobject
            (
                "transportProperties",
                runTime.constant(),
                mesh,
                IOobject::MUST_READ_IF_MODIFIED,
                IOobject::NO_WRITE
            )
        );

const dictionary& transportProps = mesh.lookupObject<IOdictionary>("transportProperties");

Info<< "Reading Kappa into kappa (permeability)\n" << endl;

const dimensionedScalar kappa = dimensionedScalar("Kappa", transportProps);

Info<< "Reading Mu into mu (viscosity)\n" << endl;

const dimensionedScalar mu = dimensionedScalar("Mu", transportProps);

Info<< "Reading gamma\n" << endl;

const dimensionedScalar gamma = dimensionedScalar("gamma", transportProps);

Info<< "Reading D (diffusivity)\n" << endl;

const dimensionedScalar D = dimensionedScalar("D", transportProps);

Info<< "Reading velocity u\n" << endl;

volVectorField u
        (
            IOobject
            (
                "u",
                runTime.timeName(),
                mesh,
                IOobject::MUST_READ,
                IOobject::AUTO_WRITE
            ),
            mesh
        );
\end{verbatim}

\section{Appendix D: Sample fvSchemes}
\label{section_fvSchemes}
\begin{verbatim}
/*--------------------------------*- C++ -*----------------------------------*\
  =========                 |
  \\      /  F ield         | OpenFOAM: The Open Source CFD Toolbox
   \\    /   O peration     | Website:  https://openfoam.org
    \\  /    A nd           | Version 7
     \\/     M anipulation  |
\*---------------------------------------------------------------------------*/
FoamFile
{
    version     2.0;
    format      ascii;
    class       dictionary;
    location    "system";
    object      fvSchemes;
}
// * * * * * * * * * * * * * * * * * * * * * * * * * * * * * * * * * * * * * //

laplacianSchemes
{
    laplacian(p)    Gauss linear corrected;
    laplacian(D,phi)    Gauss linear corrected;
}

divSchemes
{
    div(-kappa*mesh.magSf()*snGrad(p)/mu,phi)    Gauss linear;
    div(phi*gamma*(1-phi)*grad(phi)/(mag(grad(phi))+dimensionedScalar("eps",dimensionSet(0,-1,0,0,0,0,0),1e-10)))    Gauss linear;
}

gradSchemes
{
    grad(phi)    Gauss linear;
    grad(p)    Gauss linear;
}

// ************************************************************************* //
\end{verbatim}

\end{document}